\documentclass[aps,prd,preprint,showpacs,nofootinbib]{revtex4}

\usepackage[dvips]{epsfig}
\usepackage{amsfonts,amssymb,amsmath}

\usepackage{amsfonts,amssymb,amsmath}

\newcommand{\gYM}{g_{_{\text{YM}}}}

\newcommand{\be}{\begin{equation}}
\newcommand{\ee}{\end{equation}}
\newcommand{\ben}{\begin{displaymath}}
\newcommand{\een}{\end{displaymath}}
\newcommand{\bea}{\begin{eqnarray}}
\newcommand{\eea}{\end{eqnarray}}
\newcommand{\bean}{\begin{eqnarray*}}
\newcommand{\eean}{\end{eqnarray*}}
\newcommand{\nn}{\nonumber \\}

\def\l {\lambda}
\def\a {\alpha}

\def\b {\beta}
\def\g {\gamma}

\def\s {\sigma}

\def\vp {\varphi}

\renewcommand{\t}{\theta}

\newcommand{\ads}[1]{\mbox{${AdS}_{#1}$}}
\newcommand{\adss}[2]{\mbox{$AdS_{#1}\times {S}^{#2}$}}

\newcommand{\ve}{\varepsilon}
\newcommand{\tr}{\mbox{Tr}}

\renewcommand{\theequation}{\arabic{section}.\arabic{equation}}

\newcommand{\beq}{\begin{equation}}
\newcommand{\eeq}{\end{equation}}
\newcommand{\beqr}{\begin{displaymath}}
\newcommand{\eeqr}{\end{displaymath}}
\newcommand{\beqa}{\begin{eqnarray}}
\newcommand{\eeqa}{\end{eqnarray}}
\newcommand{\beqar}{\begin{eqnarray*}}
\newcommand{\eeqar}{\end{eqnarray*}}
\renewcommand{\k}{\kappa}
\newcommand{\m}{\mu}
\newcommand{\n}{\nu}

\renewcommand{\r}{\rho}

\newcommand{\cN}{{\cal N}}

\newcommand{\cO}{{\cal O}}
\newcommand{\cL}{{\cal L}}

\newcommand{\half}{\ensuremath{\frac{1}{2}}}

\newcommand{\bi}{\ensuremath{\bar{\imath}}}

\newcommand{\N}[1]{\ensuremath{\cN=#1}}

\def \ve {\varepsilon}

\newcommand{\chr}{\cosh  \r_0}
\newcommand{\shr}{\sinh  \r_0}

\def \ty {{\tilde x}}

\def \TT {{\cal T}}

\begin{document}

\title{\LARGE Spiky strings, light-like Wilson loops and pp-wave  anomaly}

\author{M. Kruczenski}
\affiliation{Department of Physics, Purdue University,
W. Lafayette, IN 47907-2036, USA}
\email{markru@purdue.edu}
\author{ A.A. Tseytlin}
\affiliation{Blackett Laboratory, Imperial College,
London SW7 2AZ, U.K. \footnote{Also at
 Lebedev  Institute, Moscow.} }
\email{tseytlin@imperial.ac.uk}

\date{\today}

\begin{abstract}
We consider rigid rotating closed strings with spikes 
 in AdS$_5$  dual to certain higher twist operators in  \N{4} SYM theory. 
 In the limit of large spin when the spikes reach the boundary of AdS$_5$, 
the solutions with  different numbers  of spikes are related by conformal transformations, 
implying that 
their energy is determined by the 
same function of the `t Hooft coupling $f(\lambda)$ that appears in 
the anomalous dimension of twist 2 operators or  in the  cusp anomaly. 
In the limit when the number of spikes goes to infinity, we find 
an equivalent description in terms of a string moving in an AdS
pp-wave background. From the boundary theory 
point of view, the corresponding description is based 
on the  gauge theory  living in a 4d   pp-wave space. 
Then,  considering a  charge moving at 
the speed of light, or a null Wilson line, we find that 
the integrated  energy  momentum tensor has a logarithmic
UV  divergence that we call the ``pp-wave anomaly''. 
The  AdS/CFT correspondence implies that, for \N{4} SYM, this  pp-wave anomaly  
should have the same value as the cusp  anomaly.  We verify this at lowest order 
in SYM  perturbation theory. As a side result of our string theory analysis,  
we  find new open string solutions in the Poincare patch of the standard AdS  space
which  end  on a light-like Wilson line and also in two parallel light-like Wilson 
lines at the boundary. 

\end{abstract}



\preprint{Imperial-TP-AT-2008-1 }

\maketitle


\overfullrule=0pt
\parskip=2pt
\parindent=12pt
\def \r {\rho}
\def \th{\theta}
\def \s {\sigma}
\def \t{\tau}
\def \ov {\over}
\def \sh {\sinh}

\def \foot {\footnote}
\def \bi{\bibitem}

\def \tr {{\rm tr}}
\def \ha {{1 \over 2}}
\def \td {\tilde}
\def \ci{\cite}
\def \no {\nonumber}

\def \ww {\Omega}
\def \const {{\rm const}}
\def \ss {\sum_{i=1}^3 }
\def \t {\tau}
\def\S{{\mathcal S} }
\def \nn {\nu}
\def \la{\label}
\def \sql {\sqrt{\lambda}}
\def \r {\rho}
\def \del {\partial}
\def\be{\begin{eqnarray}}
\def\ee{\end{eqnarray}}
\newcommand{\rf}[1]{(\ref{#1})}

\section{Introduction}

 The AdS/CFT correspondence 
 provides a concrete example of the relation between gauge theories in the
large-N limit and string theory \cite{malp}. 
In particular, \N{4} SYM theory is seen to be dual to type IIB string theory  
on \adss{5}{5}. 
In establishing this relation 
an important 
role is played by classical string 
solutions that can be mapped to  ``long''  gauge-theory 
operators  with large effective quantum numbers.
An  example is provided  by 
 strings rotating in the $S^5$ part  of
  \adss{5}{5}; 
    it improved our 
understanding of the AdS/CFT and produced numerous interesting checks of the correspondence
(see \ci{rev} for reviews).

 Another interesting case, that will  concern
  us here, is the relation between strings rotating in \ads{5} 
and twist two operators \cite{GKP} 
as well as its generalization to the relation between spiky strings and 
higher twist operators \cite{bgk,k}. 
In field theory at weak coupling  \cite{kor,make} and also, via  AdS/CFT,  at strong coupling
\cite{kru,Makeenko,krtt,am1,am2}, it can be seen that
the anomalouos dimension of twist two operators \ci{grw}
is related, for large spin,  to the cusp anomaly of a light-like 
Wilson loop. The  cusp anomaly defines a function of the `t Hooft coupling 
$f(\lambda)$, 
\be   E-S = f(\l) \ln S   \ , \ \ \ \ \ \ \ \  S \gg 1   \ , \la{EvsS}
 \ee
 which has
been studied at small coupling using SYM  perturbation theory \cite{lip} and at strong coupling 
using $AdS_5 \times S^5$ 
string sigma model   perturbation theory \cite{ft2,ftt,rt}.  
Recently, the development of the  integrability approach culminated in the  
 proposal of  an integral equation \ci{es,bes}   that describes the  function 
 $f(\lambda)$
to any order in both weak \ci{bes}  and strong \ci{kle,bas}  coupling expansion
and which passed all known perturbative tests.

 In the present paper we consider  the higher twist operators of the general form 
$\cO = \tr \left[\left(\nabla_+^{\ell} \Phi \right)^n\right]$ which have spin $S=n\ell$ and twist $n$ and which 
where argued in \cite{k} to be dual to certain rotating spiky string configurations
in $AdS_5$. In the limit of large spin, keeping the number of spikes fixed,  the 
corresponding string solutions   have  their  spikes   reaching  
the boundary of $AdS_5$ 
and are dual to Wilson loops with parallel light-like lines.

 We study these solutions concentrating on
  the ``arcs'' between the two spikes. The shape of these
   arcs is determined solely
by the angle $\Delta\theta=\frac{2\pi}{n}$ between the two spikes. Moreover, 
we show 
that the solutions corresponding to 
different values of this  angle are related to one another\footnote{This was independently 
observed by M. Abbott and I. Aniceto by embedding the spiky solutions in the sinh-Gordon model \cite{Abbott}.}  
by $SO(4,2)$ isometries of $AdS_5$. 
Since the folded rotating string of \cite{GKP} is a particular $n=2$   case 
 when  the angle between the 
spikes is $\Delta\theta=\pi$,
 this implies  that the anomalous dimensions of all the dual 
   operators should be  determined by the same 
function $f(\lambda)$.

 A  case of  
  particular interest is when the angle between the adjacent  spikes  becomes small 
($\Delta\theta\rightarrow 0$). This corresponds to the number
 of spikes $n$  going to infinity  and one  can then  concentrate    just on 
 a single arc. 
 It turns out that an   appropriate way to take this  limit is by a certain 
  rescaling of the coordinates in the  metric. 
Then
the \ads{5} metric in global coordinates reduces to a background that can be interpreted as 
  a pp-wave on top of \ads{5} in Poincare coordinates.  
  At the same time,  the boundary metric transforms from 
   $R\times S^3$ to  a 4d pp-wave, one 
  which is also conformal to 
  the  Minkowski space $R^{3,1}$. 
  
  The string solution in this limit ends on 
  two parallel light-like  lines at the  boundary.
  Computing the conserved momenta associated to such solution it follows
that the function $f(\lambda)$ is determined just by a divergence near
 the boundary and can be found  by  considering a surface ending on just  a single light-like
 line.

From the point of view of the boundary  theory, i.e.   gauge theory in the 4d pp-wave background, 
the light-like Wilson line corresponds 
to a point-like charge moving at the speed of light
in the direction $x_+$  in the  pp-wave  metric  
\beq
 ds^2 = 2dx_+ dx_- + dx_1^2 + dx_2^2 - (x_1^2+x_2^2) dx_+^2  \ . \la{cnc}
\eeq
We  should then  compute 
\beq  
\g(\l) =  -   \lim_{\ve \to 0 } \ve  \frac{\partial}{\partial\ve } \ P_+   \ ,  
\eeq
where $\epsilon \to 0$ is a UV cutoff   and $P_+$ is the 
expectation value of the null  component of the  momentum operator
in the  presence of the Wilson line 
\beq
 P_+ = \frac{1}{N} \int dx_- d^2x\  \sqrt{-g}\  \langle T^+_+\  \tr\ {\cal P} 
       \big[ e^{-i\gYM \int A^a_+ t^a dx^+}\big] \rangle|_{_{\text{pp-wave}}}  \ . \la{pyt} 
\eeq
Here $T^\m_\n$ is  the gauge theory energy-momentum   tensor.
The function $\g(\l)$ which controls the divergence of the energy-momentum near
 the charge may be called  a ``pp-wave anomaly''.
 The analysis on the  string-theory side suggests  that it should be
 related to the twist 2 anomalous dimension, or the cusp anomaly, by  
\be 
\g(\l)=\frac{1}{4}f(\l)\ . \la{hgh}
\ee
 This leads to   a novel interpretation of the cusp anomaly in the 
case of a conformal field theory. 
If the gauge theory is not conformal $\g(\l)$ is not 
necessarily related to the cusp anomaly, and 
 defines a new quantity which may  be interesting to study. 
 
\

The paper is organized as follows. 
In section 2 we shall consider  the infinite spin limit of the spiky string solution of \ci{k}
and show the  equivalence of its single-arc portion to the straight rotating string 
by  performing a conformal boost  in global coordinates of $AdS_5$. 

In section 3 we shall focus on  a special case when the arc between the two spikes gets small  and  approaches the boundary (i.e. when the number of spikes 
in the original solution   goes to infinity). 
It can be studied by first taking a certain limit of the metric of the 
global $AdS_5$ space (analogous to the  Penrose  limit)  that produces the 
$AdS_5$ metric  in Poincare coordinates with a  special pp-wave 
propagating on top of it. 

In section 4 we shall find the string  solution in this pp-wave  background 
that corresponds to the original  small-arc configuration, i.e. a world surface of an open string that ends on two parallel  null lines at the boundary of $AdS_5$. We shall 
compute the null  components  of the corresponding momentum  showing that 
$P_+$ has a logarithmic UV   divergence and that $P_+\approx \frac{\sqrt \l}{2\pi} \ln P_-$.
We shall  also consider a ``half-arc''   solution that ends on a single null line
and which is already sufficient to compute  the pp-wave anomaly coefficient
at strong coupling  and find that it is  equal to the strong-coupling value of the 
cusp anomaly \rf{hgh}.  

In section 5 we shall perform the corresponding computation in the boundary  gauge theory
and  confirm the relation \rf{hgh} also at leading order in  weak-coupling expansion.
Section 6 will contain some concluding remarks. 

In Appendix A we will show that the $AdS_5$  pp-wave   background  found in  section 3 
is still locally equivalent to Poincare patch of 
$AdS_5$ and  also demonstrate  how to construct open string solutions  in $AdS_5$ 
that end on two or one null lines at the boundary.
Appendix B will give details of the solution of the Maxwell 
equations in 4d pp-wave background with a light-like source which is used in  section 5. 
Appendix C   will contain a discussion of a generalization of the infinite spin spiky string 
 solution from section 2 to the case of a  non-zero angular momentum in $S^5$.

 
\setcounter{equation}{0}
\setcounter{subsection}{0}
\section{Infinite spin limit of the spiky string}

 Below we shall consider a 
 limit of the rotating 
 spiky string of \ci{k}, namely, the large spin limit, for fixed number of spikes, 
  in which the end-points of the spikes 
reach the boundary of the $AdS_5$. Since the spiky string is a generalization of the rotating 
folded string of \cite{GKP}, the limit is similar to the one in \ci{ft2,ftt}. In particular, 
 the limiting solution is sufficient in order 
 to reproduce the large spin limit behavior of 
the energy   ($E-S \sim \ln S$). Also, since this limiting  string touches the   
boundary, the corresponding world-surface has, as in \ci{krtt}, an open-string, 
i.e. Wilson loop,   interpretation.

\subsection{Rotating spiky string solution}

 Consider the \ads{3}  part of \ads{5}    metric 
\beq\la{ada}
 ds^2 = -\cosh^2 \r\ dt^2 + d\r^2 + \sinh^2 \r\ d\theta^2
\eeq
 and a rigidly rotating string configuration described  by the ansatz
\beq 
 t = \tau ,\ \ \ \ \ \ \theta = \omega\tau +\sigma,  \ \ \ \ \ \ \ \r=\r(\s)  \la{hp}
\eeq
The Nambu string  action and conserved quantities are given by
\beqa
 I &=& -T \int d\t d\s\  \sqrt{\r'{}^2 (\cosh^2\r -\omega^2\sinh^2\r) + \sinh^2\r\cosh^2\r}\ 
 \la{ac}
 \\ 
  P_t &=& E = T \int d\s\  \frac{\cosh^2\r \left(\r'{}^2
  +\sinh^2\r\right)}{\sqrt{\r'{}^2 (\cosh^2\r -\omega^2\sinh^2\r) + \sinh^2\r\cosh^2\r}} \\
  P_{\theta} &=& -S = -\omega T\int   d\s\
  \frac{\r'{}^2\sinh^2\r}{\sqrt{\r'{}^2 (\cosh^2\r -\omega^2\sinh^2\r) + \sinh^2\r\cosh^2\r}}\ 
\eeqa
 where $T= \frac{ \sql}{2 \pi}$ is the string tension and we defined the momenta as 
 $P_\mu = \int \frac{\partial \cL}{\partial (\partial_\tau X^\mu)} d\s$, \
 \ $I\equiv  \int d\t d\s\ \cL \equiv   T \int d\t d\s\  L$.  Here $E=P_t$ is the energy and 
$S=-P_{\theta}$  is the spin.

As   follows from the above action, the  first   integral of the
 equation for $\rho(\s)$ is \ci{k}
 \bea
  -\rho' \frac{ \del L}{\del \r'}   + L =  \frac{ \sinh^2 2\r}{\sqrt{ (2\r')^2 
  (\cosh^2\r -\omega^2\sinh^2\r)    + 
  \sinh^2 2\r }}  &  \equiv&   \sinh 2 \r_0  \la{op}   \ ,
 \eea 
  where the constant of integration $\r_0$  will be the minimal value of $\r$. 
The   maximal  value  $\r=\r_{\rm max}$   corresponds to 
$\cosh^2\r -\omega^2\sinh^2\r = 0$, i.e. 
$\coth \r_{\rm max} = \omega$. 
 The resulting 
solution describes a string with spikes. 
For large spin we obtain the relation
\beq
 E - S \simeq n T \ln S\ , 
\label{spikyE}
\eeq
where $n$ is the number of spikes and the terms we ignore are constant 
or vanishing in the $S\rightarrow \infty$ limit. This  energy-spin relation
is determined entirely by the infinite spin limit. 
This is also the  limit when   $\omega \to 1$, i.e. it    corresponds to 
the case when the ends of the spikes approach the boundary of $AdS_5$. 
In that 
limit the shape of the string simplifies as we discuss in the next subsection.

\subsection{Infinite spin limit}

 Let us consider the solution of the previous subsection in the limit when 
  the spikes touch the \ads{5} boundary.
Such solution corresponds to the value $\omega=1$ and gives the
 dominant contribution to the energy at large spin.
 Interestingly,  it  happens to have a very simple analytic form.
 As   follows from \rf{op} 
 for $\omega=1$ 
 \bea
   \frac{ 2d \r}{          \sinh 2 \r \ \sqrt{ \sinh^2 2 \r_0\  \sinh^2 2 \r -1} }  =d \s \ , 
   \eea
  so that  integrating  this equation we get 
 \be\la{htp}
\tan (\s + c)  = \frac{ \sinh 2\r_0\ \cosh 2\r}{ \sqrt{ \cosh^2 2\r - \cosh^2  2\r_0 }} \ . 
\ee
We can choose $c= \frac{\pi}{ 2}$ so that  
$\r(0)=\r_0$  and $\r(\s \to \pm \s_0)  \to \infty$. Then 
\be 
\coth 2\rho = {\cos\s \ov \cos \s_0}   \ , \la{kol}\\
\cot \s_0 = \sinh 2\r_0  \ .  \la{kp} 
\ee
Equivalently, 
\beqa
\cosh 2\r &=& 
 \frac{\cos\s}{\sqrt{\sin^2\s_0 - \sin^2\s}}\ , \no \\
\sinh 2\r &=& 
  \frac{\cos\s_0}{\sqrt{\sin^2\s_0 - \sin^2\s}}  \ . \la{roo}
\eeqa
 Since  $\rho\rightarrow  \infty$ when $\s \rightarrow \pm \s_0$ we have 
  $\Delta \theta= 2\s_0 = \frac{2\pi}{n}$ where $n$ is the number
of spikes. 
In general, the spiky string is a function of two parameters,
 the angular momentum and the number of spikes or,  equivalently, 
$\rho_{_{\rm {min}}}$, and $\rho_{_{\rm {max}}}$ -- the radii at the valleys and the spikes.
 Since we took $\rho_{_{\rm {max}}}\rightarrow\infty$, keeping $\rho_{_{\rm {min}}}$ fixed,  only
$\rho_{_{\rm {min}}}$ remains as a parameter, or equivalently the number of spikes given here by $\s_0$.

 Since  in this limit 
  the spikes touch the boundary, we
can now take only one arc between the two spikes and ignore the rest of the string. 
In the open-string picture 
\ci{krtt}, this arc 
  corresponds to   a Wilson loop surface ending on 
two parallel light-like lines  at  the boundary  (the spikes
move at the speed of light). 

Let us rewrite   the above solution in the global  embedding coordinates.
First, note that 
\beq
\cos\s =  \cos(\theta-t) = \cos\theta\cos t +\sin\theta\sin t = \cos\s_0 \coth2\rho
\eeq
Now, if we  use the  parameterization  (with $Y_3=Y_4=0$)
\beq\la{hqp}
 Y_1 = \sinh\rho\ \cos\theta, \ \ Y_2 = \sinh\rho\ \sin\theta, \ 
 \ Y_5 = \cosh\rho\ \cos t, \ \ Y_6=\cosh \rho\ \sin t \
\eeq
then 
\beqa
 Y_1 Y_5 + Y_2 Y_6 &=& \cos\s_0 \sinh\rho\cosh\rho\coth2\rho\no \\
                   &=& \half \cos\s_0\ \left(Y_1^2+Y_2^2 + Y_5^2+Y_6^2\right)\no \\ 
                   &=&  \cos\s_0\ \big(Y_5^2+Y_6^2 -\half \big)\ . \la{hhp}
\eeqa
 We see  that the  world  surface as a 2d surface in 
 $R^{4,2}$
 is determined by the following system of 4 equations
\beqa
&& Y_1^2 +Y_2^2 + Y_3^2 + Y_4^2 - Y_5^2 - Y_6^2 = -1\ ,  \no\\
&& Y_1 Y_5 + Y_2 Y_6 - \half \cos\s_0 \left(Y_1^2+Y_2^2+Y_5^2+Y_6^2\right) = 0\ ,  \no\\
&& Y_3=Y_4=0\ .   \la{li}
\eeqa
Using the $SO(4,2)$ symmetry of \ads{5} we can put the second 
quadratic equation into a  simple  form, thus finding 
\beqa
&& Z_1^2 +Z_2^2 + Z_3^2 + Z_4^2 - Z_5^2 - Z_6^2 = -1\ , \no \\
&& Z_1 Z_5 - Z_2 Z_6 = 0 \ , \la{uni} \\
&& Z_3=Z_4=0\ . \no
\eeqa
Here   $Z_m$  are  defined by 
\beqa
 Y_1 &=& Z_1 \chr + Z_5 \shr  \no\\
 Y_5 &=& Z_5 \chr + Z_1 \shr  \no \\
 Y_2 &=& -Z_2 \chr + Z_6 \shr \no \\
 Y_6 &=& Z_6 \chr - Z_2 \shr  \la{ki} \ . 
\eeqa
We used \rf{kp}, i.e.  that 
$\tanh 2\r_0 = \cos\s_0$. 

We see therefore that all solutions that we found in this limit, parameterized by $\s_0$,
are  actually 
related to one another by the two $SO(4,2)$ boosts in the 1-5 and 2-6 hyperbolic planes. 
Namely,   a generic arc of the spiky string connecting
two spikes  reaching the boundary 
is  related by these  $SO(2,4)$ boosts  to the infinite spin limit \ci{ft2,ftt}
 of the straight  folded rotating string of \cite{GKP} 
 which 
in our present notation 
 corresponds to $\s_0=\frac{\pi}{2}$   or  $\r_0=0$  (the center  
 of the string is at the center of $AdS_3$). 
 
 As follows from \rf{uni},  this  surface   can be  parametrized simply by\foot{ 
 This corresponds to the choice of 
  conformal gauge   \ci{krtt}. For simplicity, we use the same
 notation  $\s$ and $\t$  for the conformal-gauge  world-sheet coordinates.
 These are rescaled coordinates: the original ones  in conformal gauge 
 should contain the scale factor  $\kappa \approx { 1\ov  \pi } 
 \ln S \gg 1$ (related to $\omega \to 1$), i.e.  while the original 
 $\s$ varies  between $\pm {\pi \ov 2}$, the rescaled one 
 varies between $\k \to \pm \infty$. The same remark will apply to $(\s,\tau)$
 in $Y_m$ below. 
 } 
  \bea 
 Z_1&=& \sinh \sigma\  \cos \tau  , \ \ \ \ \ \ \ \
 Z_2= \sinh \sigma\  \sin \tau  ,\no \\ 
 Z_5&=& \cosh \sigma\  \sin \tau  , \ \ \ \ \ \ \ \ 
 Z_6= \cosh \sigma\  \cos \tau  . \la{jp}
 \eea
Then  we get from \rf{ki}
\beqa
Y_1 &=& \sinh \sigma\  \cos \tau\ \cosh \r_0 +\ \cosh \sigma\  \sin \tau \ \sinh \r_0 \no\\
Y_2 &=& -\sinh \sigma\  \sin \tau\ \cosh \r_0 +\ \cosh \sigma\  \cos \tau \  \sinh \r_0 \no\\
Y_5 &=&  \cosh \sigma\  \sin \tau \ \cosh \r_0 +\  \sinh \sigma\  \cos \tau\ \sinh \r_0\no 
\\
Y_6 &=& \cosh \sigma\  \cos \tau\  \cosh \r_0  - \ \sinh \sigma\  \sin \tau \ \sinh \r_0  \la{jo}
\eeqa
 so that  the induced 
 metric is $ds^2 = -d\tau^2 + d\sigma^2$.
  This  determines the expression 
 for our  solution (i.e. $t=\tau,\ \ \theta= \tau + \sigma$ and  $\rho(\s)$ 
 given by \rf{roo})    when  it is transformed to conformal gauge.
  
 As was argued in 
 \cite{k},  the spiky string state corresponds to  higher
 twist operators with   maximal anomalous dimension.
  The above discussion  shows that, 
 for large spin,  the anomalous dimension of
all such  operators is determined by the same universal scaling 
function  $f(\lambda)$ that appears for  twist two operators. This is simply
because the corresponding 
surfaces are related by conformal boosts and thus 
the associated string 
 partition functions should be the same  to   all loop  orders (see \ci{krtt}).


\setcounter{equation}{0}
\setcounter{subsection}{0}

\section{``Near-boundary string''  limit}

The solution \rf{kol} found in the $S\rightarrow\infty$ limit   admits two special cases.
 One is   when  the variable $\s_0$ which determines the angular distance 
between the two spikes  ($\Delta\theta=2\s_0$)   approaches 
 $ {\pi \ov 2}$, i.e  $\ha \Delta\theta=\s_0\to {\pi\ov 2} $. Then  $\r_0 \to 0$ 
 and  the arc becomes  the straight string ($\Delta \theta= \pi$)
 passing through the center of $AdS_5$ 
 with its ends reaching the boundary. In this limit the solution
 ($ \coth 2 \r= {1 \ov {\pi \ov 2} - \s_0}  \cos \s $)
  looks   singular, implying that  some rescaling is to be made. 
 
 Another special case corresponds 
 to $\s_0 \to 0$, i.e. $\r_0 \to \infty$, $\Delta \theta\to 0$
 when   the arc between the two adjacent  spikes  becomes  small and  represents 
 an open fast-rotating 
 string  located close to  the boundary  with its ends moving along null lines at 
  the boundary. 
 According to \rf{ki},  this case corresponds to an infinite boost 
 of the  straight string passing through the center.\foot{In this limit 
 $\Delta \theta = { 2 \pi \ov  n} = 2 \s_0 \to 0$  
 so that the total number of spikes of the original closed string goes to infinity.
 While each arc still contributes $ T \ln S$ to the energy, 
 the  total energy \rf{spikyE} of the closed string then becomes infinite.}

 In global coordinates 
 the limit
 $\r_0 \to 0$ corresponds to no  boosts in  \rf{ki}  so that 
 $Y_m=Z_m$  in \rf{jp}.  In the limit $\r_0 \to \infty$ we get from  \rf{ki}
 \beqa
&& Y_1 - Y_5 \to 0, \ \ \   Y_1 + Y_5    \to   e^{ \r_0} (  Z_1 + Z_5) \ ,\no \\
&& Y_2-  Y_6 \to 0, \ \ \ \  Y_2 +   Y_6   \to   e^{ \r_0}  (-Z_2 + Z_6)  \la{kdi} \ ,  
\eeqa
 i.e. this   corresponds to an infinite boost in the two (1,5),(2,6) planes.

  In that second case when $\s_0$  and thus $\s$  are  small while 
 $\r$  is large   we can {\it approximate} the exact solution 
 \rf{hp},\rf{roo} as: 
\beqa
 t &=& \tau\no \ , \ \ \ \ \ \ 
 \theta = \s + \tau\no \ , \\
 \half e^{2\rho} &\simeq& \frac{1}{\sqrt{\s_0^2 - \s^2}}\ .   \la{ba}
\eeqa
 In this limit  the ends of the string   follow  two light-like lines at the boundary
 which are very close to each other. 
 

 Since for $\s_0 \to 0$  the string is  located  close to the boundary 
  we  may  
 expect that we can ignore the periodicity in $\theta$ and get the corresponding 
 solution in the Poincare patch  $ds^2 = { 1 \ov z^2} ( - dt^2 + dx^2 + dz^2)$ 
 by identifying
\beqa
 t \to  t \ , \ \ \ \ \ \ 
 \theta  \to  x \ , \ \ \ \ \ \ 
   \half e^{-\rho} \to z  \ . 
\eeqa
The  solution \rf{ba}   then reads as 
\beqa
 t &=& \tau  \ , \ \ \ \ \ \ \ \ 
 x = \s + \tau\no \\
 z &=& \frac{1}{2\sqrt{2}} \left(\s_0^2 - \s^2\right)^{\frac{1}{4}}
\label{Zsol}
\eeqa
 ending at the boundary $z=0$ on  the two parallel null lines 
  $x-t \equiv x_- = \pm\s_0$. 
  
However,  \rf{Zsol}   is not an exact 
solution in the Poincare patch. 
   Still,  as we shall now show, 
    one  can take a particular (infinite-boost) limit of 
the global \ads{5} metric and obtain a new metric
for which 
 eq.(\ref{Zsol}) will be  an exact  solution.

 Let us  start by  writing the global \ads{5} metric as
\beqa
 ds^2 &=& -\cosh^2\r\ dt^2 + d\r^2 + \sinh\r^2\ d\Omega_{[3]}^2 \no \\
      &=& -\cosh^2\r\ dt^2 + d\r^2 
         + \sinh\r^2\  \frac{(1-\frac{x_i^2}{4})^2\  d\theta^2 + dx_i dx_i }
	 {(1+\frac{x_i^2}{4})^2}   \ , \la{add}
\eeqa
 where $x_{1},x_2$  and  $\theta$ parameterize the 3-sphere. 
 We can now make a change of coordinates $(\r,t,\theta) \to (z, x_+, x_-)$ \foot{In what follows we
 shall use a formal  notation in which $x_+\equiv x^+, \ x_- \equiv x^-$.}
\beqa
 \r = -\ln(2z)    \ , \ \ \ \ \ \ \  
 t = x_+ - x_-  \ , \ \ \ \ \ \ 
 \theta = x_+ + x_- \ ,  \la{si}
\eeqa
and  a rescaling  by a parameter $\ve$
\beqa
 x_+ &\rightarrow&\ \m^{-1}\ x_+\no \\
 x_- &\rightarrow&\ 8 \m\ \ve^2\ x_- \no\\
 x_i &\rightarrow&\ 4 \ve\ x_i \no\\
 z &\rightarrow&\ \ve\ z  \la{zc}
\eeqa
Here we also introduced a (spurious) mass scale  $\m$.
In the  limit when $\ve\to 0$  we get $\theta \to t$   and $\r \to \infty$
while $x_i \to 0$, i.e. this limit  focuses on   a small  fast-rotating 
 string   located near the boundary  and near the origin of 
the transverse $x_i$ space. The end-points of the string 
(which are then close to its center of mass)  follow the  massless geodesic 
$t=\tau, \ \theta= \tau \pm \s_0 ,  \ x_i= 0, \ \r= \infty\   (z=0) $.

Taking  the limit $\ve \rightarrow 0$  in  \rf{add},\rf{zc} 
 and keeping
only  the leading $\ve$-independent  terms   we get 
the metric which such small string  ``sees'':\foot{This limit is similar to the Penrose limit
used in the case of the $S^5$ geodesic \ci{pen}
but notice that here we do not rescale 
the overall coefficient of the   metric, i.e. the string tension. 
In fact  it appears to be a special case of the
``conformal Penrose limit''  considered in \ci{guven}.}
\beq
ds^2 = \frac{1}{z^2} \left[ 
2 dx_+ dx_- - \m^2 \left(z^2 + x_i^2\right) dx_+^2 + dx_i dx_i + dz^2 \right]
\label{ppbkg}
\eeq
This metric   may be interpreted as  a pp-wave
 in the  \ads{5} space.\foot{For a discussion  of 
 various ``pp-wave on top of   $AdS_n$''  solutions see 
 \ci{cv,cha}.} 
 Note  that the mass parameter $\m$  can be set 
 to 1  by a boost in the $x_+,x_-$ plane.

 According to the 
AdS/CFT duality,   the corresponding string theory 
should be dual to the \N{4} SYM gauge theory  defined 
 on the 4d  pp-wave background\foot{For  the 
 structure of the action of  this gauge theory
  see \ci{mee}; see also \ci{qqj} for some studies of
  the  quantum gauge  theory in pp-wave backgrounds.} 
\beq
ds^2 =
2 dx_+ dx_- - \m^2 x_i^2 dx_+^2 + dx_i dx_i \ . 
\label{ppbdy}
\eeq
This does not contradict the fact that 
\rf{ppbkg} 
was obtained as a limit of the ``empty'' $AdS_5$ space --
 the metric \rf{ppbdy}  is conformally-flat (the 5d metric \rf{ppbkg}
 is, in fact, locally equivalent to the $AdS_5$  metric  \ci{cha}, see Appendix A).

Starting with \rf{ppbkg} we can 
  now  consider two parallel light-like lines in the
 $x_+$ direction and check (see below)  that
there is  indeed a solution of the form of \rf{Zsol}, i.e. 
 $z=z(x_-)$,  ending on them.


\setcounter{equation}{0}
\setcounter{subsection}{0}
\section{Cusp anomaly from null Wilson line in a pp-wave: strong coupling}


Suppose we 
 consider the planar \N{4} SYM theory 
 in the pp-wave metric (\ref{ppbdy}) and compute the expectation value of the Wilson loop 
 bounded by two infinite   parallel  light-like lines in the  direction $x_+$:
 \ \ \  $x_-=\pm \s_0, \ 
 x_i=0$, \ $\s_0$-const. 
 Then according to AdS/CFT, 
 the strong `t Hooft coupling limit of  the expectation value of
 such Wilson loop should be determined \ci{maldrey}
  by the string action  for  the  minimal-area surface in the pp-wave 
metric (\ref{ppbkg}) ending on these two null  lines. 

Let us find the corresponding  string solution  by assuming that 
it  has  the form 
\beq\la{bas}
 x_+ = \tau 
 \  ,\ \ \ \ x_-=\sigma,\ \ \ \ z=z(\sigma),\ \ \ \ x_i=0\ , \ \ \ \ \ 
 -\s_0 \leq  \s \leq \s_0 \ . 
\eeq
 Then the string  action corresponding to \rf{ppbkg}  
  becomes\foot{One can check directly that the above ansatz is indeed
 consistent with all the relevant equations.} 
\beq
I= -T \int  d\tau d\sigma\ \frac{1}{z^2} \sqrt{1+  \m^2 z^2 z'{}^2}
\eeq
 Minimizing with respect to $z(\sigma)$ we obtain 
(we set $\m=1$  in what follows)
\beq
 z' = \frac{1}{z}
 \sqrt{\frac{z_0^4}{z^4}-1}
\label{zp}
\eeq
or
\beq
 z = \sqrt 2 (\s_0^2-\s^2)^{\frac{1}{4}} \ , \ \ \ \ \ \ \
  \ \ \ z_0 \equiv z_{\rm max} = \sqrt{2 \s_0} \ . \label{zpp}
\eeq
For  $\s$ changing   from $-\s_0$  to $+\s_0$  we have  $z$
tracing  both  halves  of the ``arc'' -- from 0 to $z_0$   and then 
from $z_0$ to 0. 

This solution agrees with eq.(\ref{Zsol}) after a rescaling of $\s$
(as  
follows from comparing the $x_\pm$ part of the solution, cf. \rf{Zsol} and  \rf{si}),  
  and this  is again just a limit 
of the solution we have found in  section 2.

 Let us now compute the conserved quantities\foot{Here 
 $ \cL= - T  \sqrt{- g}$, where $g$ is the induced metric corresponding to \rf{ppbkg}.
 In general, if $x_+ = \kappa \tau, \ \ x_- =  \alpha  \sigma + \beta \tau$
 then 
  $ \sqrt{- g}= z^{-2}\sqrt{ \k^2 \a^2  + (\k^2 \mu^2 z^2 - 2 \k \b) z'^2 }$.} 
\beqa
 P_+ &=& -\int^{\s_0}_{-\s_0} d\sigma\ \frac{\partial \cL}{\partial(\partial_\tau x_+)} =
  T \int^{\s_0}_{-\s_0} d\s\ \frac{1}{z^2}\sqrt{1 +  z^2  z'^2 } \ , \la{yd}\\
 P_- &=&  \int^{\s_0}_{-\s_0} d\sigma\ \frac{\partial \cL}{\partial(\partial_\tau x_-)} = 
   T \int^{\s_0}_{-\s_0} d\s\ \frac{z'^2}{z^2\sqrt{  1 +  z^2  z'^2     } } \ . \la{yyd}
\eeqa
Using eq.(\ref{zp}) we can convert the integrals over $\s$ into the integrals over  $z$. 
The latter will be   divergent at $z=0$ so we will insert a cut-off, 
$z=\epsilon\to 0$. Thus 
\beqa
 P_+ &=&  2 T \int_\epsilon^{z_0} \frac{ dz}{z \sqrt{ 1- \frac{z^4}{z^4_0} }}
  =   T \ln {{2}z^2_0\ov \epsilon^2} 
  + \cO(\epsilon) \ , \la{yye} \\
 P_- &=&  2 {T} \int_\epsilon^{z_0} {dz \ov z^3}  \sqrt{1 - \frac{z^4}{z^4_0}}
  =- \frac{T}{\epsilon^2} + \frac{\pi T}{2z_0^2} + \cO(\epsilon) \ , \la{eyy}
\eeqa
where the factor of 2 reflects  the fact that $(-\s_0,\s_0)$ 
covers $(\epsilon,z_0)$ twice.
Thus solving for $\epsilon$   we get 
\be\la{rell}
P_+ \approx   T\ln |P_-| \ . 
\ee
This can be related to the spiky string expression 
\rf{spikyE} if  we formally identify (cf. \rf{si}) 
\beq
 P_+ = P_t + P_\theta=E-S ,\ \ \ \  \ \ \ \ P_- = -P_t + P_\theta = -E-S \la{asd}
\eeq
Then, at leading order in $\epsilon\rightarrow 0$, 
\beq
 E-S \approx  T \ln { 1 \ov \epsilon^2}\ ,\ \ \ \  \ \ \ 
 \ E+S \approx     \frac{T}{\epsilon^2}  \ , 
\eeq
and since,  to  leading order, $E\approx  S$, we get
\beq 
E-S \approx  T\ln S  \ . \la{coom}
\eeq
This   agrees with  (\ref{spikyE}) since here 
 we are  considering  one arc  ($n=1$) of the full $n$-spike closed string. 

We conclude   that, if we are interested in the strong-coupling limit
of anomalous the dimension of 
 operators with  large spin, it suffices to consider  this particular string  solution 
in the \ads{5} pp-wave background.
This is analogous to the ``open-string'' computation 
of this anomalous dimension from the null cusp surface 
in the Poincare patch in \ci{kru}; the two world-sheet surfaces 
are indeed related by an analytic continuation \ci{krtt} and 
a global-coordinate boost \rf{kdi}.

\

It is useful to notice   that the above expressions \rf{yye}  and \rf{eyy}
were  essentially determined by the contribution near  $z\sim \epsilon$, 
i.e. the  result \rf{rell} 
follows from the properties of the solution \rf{zpp}  near the boundary. 
For that reason we may repeat the above discussion 
for  a surface ending not on two  but just on one null line; the  role of $z_0$
will then be played by an explicit IR cutoff $z \leq  z_{\rm max}\equiv  R$.

Indeed, let us  consider a world-line of a single massless ``quark'' at $x_-=0$.
 The corresponding  ``straight-string''  world  surface 
ending on this   world line $ x_+ = \tau$ is then 
(both in the standard $AdS_5$  and in the $AdS_5$  pp-wave   background \rf{ppbkg}) 
\beq
 x_+ = \tau,\ \ \ \  \ \ \ \ z = \sigma ,\ \ \ \ \ \ \ \ \ \ x_-=x_i=0 \ . 
\label{sWL}
\eeq
Computing the associated momenta as in  \rf{yd},\rf{yyd} or \rf{yye},\rf{eyy}
for the case of the {\it pp-wave   background}
 we shall cut
 off the integrals over $z$  at $z=\epsilon\to
0 $ and
  at $z=L\to \infty$:
\beqa\la{jqp}
 P_+ =   T \int^{L}_{\epsilon} dz \frac{\sqrt{\mu^2 z^2  } }{z^2}
      \approx { \mu  T\ov 2} \ln \frac{L^2}{\epsilon^2}\ , \\ 
 P_- =    T \int^{\infty}_{\epsilon} dz \frac{1 }{z^2 \sqrt{\mu^2 z^2  }   }     
            \approx   - \frac{T}{2\mu \epsilon^2} \ . \la{uur}
\eeqa
Here we restored  the dependence on the pp-wave scale $\mu$
to indicate that  a non-zero  value of 
$P_+$   is found only in the pp-wave case, i.e. if
 $\m\not=0$. 

These expressions are  one half smaller than  in \rf{yye}  and \rf{eyy} 
since here we effectively have only ``half'' of the previous  world surface that was ending on two null lines. 
 As   a result, 
\beq
P_+ \approx  \frac{T}{2} \ln |P_- | =  {\sql \ov 4 \pi} \ln |P_- |   \ , \la{kyy}
\eeq
or, using  \rf{asd}, 
\beq
 E-S =\frac{\sql }{4\pi } \ln S   \ . \la{yyy}
\eeq
Again, this is one half  smaller  compared to   \rf{coom} 
 due to the fact that here we had 
  only one null   line instead of two.

  The conclusion is that   in order to compute the scaling 
 function  that multiplies $\ln S$  in the 
 anomalous dimension
    it is sufficient 
  to  find  the
UV ($z \to 0$) divergence 
 of the momenta   corresponding   to the surface  in the pp-wave metric \rf{ppbkg}  ending on  the 
null  Wilson line, i.e. on 
a single  line  in the $x_+$ direction.

 This ``elementary'' or  ``half-arc'' solution thus reproduces  1/2 of 
 the anomaly coefficient   of the  two null line 
 surface  or the the one-spike solution, and thus 1/4  of the  
 anomaly  captured by straight folded rotating string (or two-spike solution).

As we shall  discuss below, one can do also  a similar   computation 
for a null Wilson line 
in the  weakly  coupled  gauge theory defined  in the
 corresponding 4d pp-wave background \rf{ppbdy}.
In the conformal gauge theory, the single  null  Wilson line  computation 
 will give the same result  as the cusp anomaly in the fundamental representation, 
 which is   1/4  smaller than  the twist 2 anomalous dimension, i.e. the 
 dimension of the operator
 like $\tr( \Phi D^S_+  \Phi)$   dual to  the  closed  folded rotating string.


\setcounter{equation}{0}
\setcounter{subsection}{0}

\section{Cusp anomaly from null Wilson line in a pp-wave: weak coupling}

As we have  found  in the previous section, 
an alternative way to compute the strong-coupling limit 
of the twist 2 anomalous dimension 
(or, equivalently, the cusp anomalous dimension)
  is to consider an open-string world surface 
in the ``$AdS_5$ plus  pp-wave'' background \rf{ppbkg} 
that ends on a  null line  at the boundary of $AdS_5$. 

This suggests that one should be able to find the  weak-coupling limit of 
the twist 2 anomalous dimension 
by considering a similar   set-up in the boundary theory 
-- the  SYM  gauge theory in the pp-wave background \rf{ppbdy}. Namely, 
we should  study  a field  produced  by a  single charge
moving at the speed of light along the $x_+$ direction 
in the 4d  pp-wave background. 
More explicitly, 
we would   like to  reproduce   the relation   \rf{kyy} at weak coupling, i.e. 
 \beq
P_+ =   \g(\l)  \ln |P_- |   \ , \la{jyy}
\eeq
where $\g$   (that we may call a ``pp-wave anomaly'' coefficient) 
in the conformal gauge theory case\foot{To the leading order in $\l$ 
that we will consider below  the distinction 
between the conformal and non-conformal cases will not be visible.}
 will turn out to be  proportional to 
 the twist 2 anomalous  dimension $f(\l)$ 
 \be   \g(\l) = { 1 \ov 4}   f(\l) \ .  \la{prt}
 \ee
  The latter  has the well-known perturbative expansion 
\ci{grw}
\be
f(\l)  = { \l \ov 2 \pi^2} + O(\l^2)   \ . \la{ff}
\ee
The scaling function $f(\l)$ is proportional to the cusp   anomalous dimension, 
i.e. it can be found  also 
as a singular part of the  expectation value of the Wilson loop 
with a  cusp formed by two null lines in  flat (Minkowski) space 
\ci{kor}. At the lowest order in gauge coupling and in the 
planar limit ($\lambda=\gYM^2 N$, \  $N \to \infty$)
the cusp  anomaly (in the fundamental representation) is
\beq\la{jqs}
 {\Gamma}_{\mbox{cusp}} =  \frac{\lambda}{8\pi^2} + O(\l^2)  =  { 1 \ov 4} f(\l)  \ . 
\eeq
Our aim will be  to check 
the validity of \rf{prt}, i.e. to reproduce \rf{ff}  or \rf{jqs} in  the 
``gauge theory in pp-wave'' set-up.

The null Wilson line along $x_+$ is BPS  (both in flat space and in the pp-wave case), 
so the corresponding expectation value is trivial, 
 $ \langle W \rangle =1$. Instead, we are  to find 
the logarithmic UV anomaly in the  light-cone energy $P_+$
in the presence of a null line or  the relation \rf{jyy} between $P_+$ and $P_-$. 
Let us define 
\be
   P_\pm  &\equiv&
   \int dx_- d^2x_i \sqrt{-g}\  \langle\ T^+_\pm\ W\  \rangle_{_{\text{pp-wave}}}  \ , 
   \la{jdq}
     \\
   \la{jfq}
  W &=& \frac{1}{N} \tr\ {\cal P} e^{-i\gYM \int A^a_+ t^a dx^+}  \ . 
\ee
Here $T^+_\pm$  are  the components of the gauge theory energy-momentum tensor  
and the expectation value   is computed in the gauge theory 
defined in the pp-wave background \rf{ppbdy}.
$t^a$ are the 
$SU(N)$  generators in the fundamental representation 
normalized as $\tr (t^a t^b) = \half \delta^{ab}$  ($a,b=1,2,...,N^2-1$). 

As we shall see, $P_+$ will be  logarithmically UV  divergent
and thus we  may define the ``pp-wave anomaly'' as 
\beq  \la{fef}
\g(\l) =  - \half  \lim_{\ve \to 0 } \ve  \frac{\partial}{\partial\ve } \ P_+   \ ,  
\eeq
where $\ve \to 0$ is a UV cutoff and the factor of two is due to the fact that $P_-$ is 
quadratically divergent in the cut-off. Then  \rf{jyy} will follow from a scaling argument described below. 

At lowest order in the gauge coupling the  expectation value \rf{jdq}
is given simply by the one-gluon exchange, i.e. by the gaussian path integral 
saturated by the classical  gauge field configuration with a source provided
 by the null Wilson line. 
 We  can then simply replace  the gauge field 
by  an abelian one $A_\mu$  including the  factor ${C_F}=\frac{N^2-1}{2}$
from the trace in  the   final expression.
 The  corresponding abelian  action is then\foot{Recall that in our notation $x^+\equiv x_+$.}
\beq\la{jqh}
 {\rm S}  = -\frac{1}{4} \int d^4 x \sqrt{-g}\ F^{\mu\nu}F_{\mu\nu} - \gYM \int A_+ dx^+ \ , 
 \ \ \ \ \ \ \ \ \ \ \ F_{\mu\nu}= \partial_\mu A_\nu - 
\partial_{\nu} A_\mu \ . 
\eeq
Notice  that for the pp-wave   in  \rf{ppbdy}  we have   $\sqrt{-g}=1$
so we will ignore this factor in   what follows. 

We are then to solve  the equations of motion for \rf{jqh}  ($i=1,2$) 
\beqa
 \partial_\mu  F^{\mu+}
  = \ \gYM\ \delta(x_-)\ \delta^{(2)}(x_i) \  ,  \ \ \ \ \ \ \ \ \ 
 \partial_\mu 
 F^{\mu -}
 = \partial_\mu 
 F^{\mu i}
 =0\la{jlq}
\eeqa
and evaluate \rf{jdq} on the solution. 
As shown in Appendix B,   the relevant solution is ($r\equiv \sqrt{ x_i x_i}$) 
\beqa
 A_- &=& \frac{\gYM}{4\pi^2 x_-} \arctan \frac{\m r^2}{2x_-} \ ,  \ \ \ \ \ \ \ \ \ \ \   A_i=0 \ , \no  \\ 
 A_+ &=& -\frac{\gYM}{4\pi^2} \ln\left(\mu^2 r^4+ 4 x_-^2\right) \ , \la{bjq}
\eeqa
i.e. 
\beqa
 F^{+-} &=& F_{-+} = -\frac{2\gYM}{\pi^2} \frac{\mu x_-}{\mu^2 r^4 + 4x_-^2}\ , 
 \no \\
 F_{+i} &=& \frac{\gYM}{\pi^2} \frac{\mu^3 r^2 x_i}{\mu^2 r^4 + 4x_-^2}\  ,
 \ \ \ \  \ \    F^{-i} = F_{+i}  +  \mu^2 r^2 F_{-i} =0 \ , \no \\ 
 F^{+i} &=& F_{-i} = -\frac{\gYM}{\pi^2} \frac{\mu x_i}{\mu^2 r^4 + 4x_-^2}  
 \ .   \la{jssq}
\eeqa
Computing  the energy-momentum tensor 
\beq
T^{\mu}{}_\nu = F^{\mu\alpha} F_{\alpha\nu} - \frac{1}{4} \delta^{\mu}{}_\nu \,
 F^{\alpha\beta} F_{\beta\alpha} \ ,  \la{jnq}
\eeq
we get 
\beqa
 T^{+}{}_{+} = \frac{\gYM^2 }{2 \pi^4} \frac{\mu^2 }{ \mu^2 r^4 + 4x_-^2}\ , 
  \ \ \ \ \ \ \ \ \
 T^{+}{}_{-} = -\frac{ \gYM^2 }{\pi^4} \frac{\mu^2 r^2}{(\mu^2 r^4 + 4x_-^2)^2}
 \la{nnjq} \ . \la{ttt} 
\eeqa
The  relevant  components of the momentum 
 (including the non-abelian  group-theory factor which in the planar limit 
 is 
  $\frac{C_F}{N}\approx {N \ov 2} $, see \rf{jdq}) are 
\beq
 P_+ = \frac{N}{2}  \int^\infty_{-\infty} dx_- d^2x  
\  T^+{}_+\ , 
\ \ \ \ \ \ \ P_- = \frac{N}{2}  \int^\infty_{-\infty} dx_- d^2 x  \  
 T^+{}_-  \ .      \la{jmq}
\eeq
 $P_+$ will be  logarithmically   divergent both in the UV and in the IR, 
while $P_-$  will be quadratically UV divergent. To extract this divergence it is
sufficient to introduce an UV regularization in \rf{ttt} by 
\be 
\mu^2 r^4 + 4x_-^2 \  \rightarrow  \  \mu^2 r^4 + 4x_-^2 + \mu^2  \ve^4\ , \ \ \ \ \ \ 
\ve \to 0   \ . \ee
Then doing the integrals in  \rf{jmq}  we get 
\beqa
P_+ &=& \frac{\lambda}{4\pi^4} 2 \pi \int_{-\infty}^{\infty} 
dx_- \int_0^\infty  d r \ r \  \frac{  \m^2 }{\mu^2 r^4 + 4x_-^2 + \mu^2  \ve^4} 
\approx  { \lambda \ov 8 \pi^2}\ \mu \ \ln  { L^2 \ov \ve^2 }  \ , \la{ppm}    \\
P_- &=&- \frac{\lambda}{2\pi^4} 2 \pi   \int_{-\infty}^{\infty} 
dx_- \int_0^\infty d r \ r \  \frac{  \m^2\  r^2}{(\mu^2 r^4 + 4x_-^2 + \mu^2  \ve^4)^2} 
\approx  - { \lambda \ov 8 \pi^2 \mu \ve^2 }  \ ,  \la{ggg} 
\eeqa
where $L\to \infty$ is an infrared cutoff  ($\mu^2 r^4 + 4x_-^2 < \mu^2 L^4$). 

Note that, with $\l$ being  dimensionless,  
the factor  of $\m$ in $P_+$ is necessary 
in order for $P_+$ to have the right (mass) dimension. 
That means that  $P_+$  
 vanishes in the flat-space limit, 
i.e. the pp-wave background   is essential  for  obtaining this
logarithmically    divergent  result.

With   $P_+$  containing  the logarithmic  UV divergence  we
 can read off the corresponding anomaly 
coefficient as in \rf{fef}. 
Equivalently,  since $|P_-| \sim \ve^{-2} \to \infty$, 
 we find, 
 fixing  the pp-wave scale $\m=1$, 
\beq
  P_+ \approx  \frac{\lambda}{8\pi^2}    \ \ln |P_-| \ .
  \la{igg} 
\eeq
This  is in  perfect   agreement  with the equations \rf{jyy},\rf{prt},\rf{ff}, 
namely with the 1-loop value of the  twist two anomalous dimension or the cusp 
anomalous dimension. 

That $P_+ $ should be proportional to $ \ln |P_-|$  can be understood on general grounds
and thus the  relation \rf{jyy}  should be true to all orders in the expansion in $\l$.
This follows from the following  scaling argument.  
The pp-wave metric (\ref{ppbdy}) is invariant, 
up to an overall scale, under the following   transformation
\beq
  \tilde{x}_+ = x_+\ ,\  \ \ \ \ \ \ \ 
   \tilde{x}_- = \xi^2 x_-\ , \ \ \ \ \ \ 
    \ \tilde{x}_i = \xi x_i \ , \la{loi}
\eeq
which is the counterpart of   the scaling symmetry in flat space
(there is a conformal Killing vector associated to \rf{loi}).
Since the \N{4} SYM theory is conformally invariant, this 
symmetry should be present in the quantum gauge theory. 
Since $x_+$ and $x_-$ scale differently in \rf{loi}, 
the corresponding  components  of the momentum $P_+$ and $P_-$ 
will have different scaling weights: zero for $P_+$  and -2 for $P_-$. 
This  implies that $P_+$ should be a dimensionless function 
of $P_- L^2$ where $L$ is an IR cutoff. 
Assuming the corresponding quantum state does not  break  the scaling symmetry spontaneously
 (in particular, that  the classical background does not contain extra mass parameters), 
 this  function will  be the logarithm.
 
In more detail, the \N{4} SYM theory defined on any conformally flat background is 
invariant  under  both  diffeormorphisms and the Weyl symmetry, and, in particular, 
under their combination that  preserves the form of the  background
 metric (see, e.g., \ci{cft}).\foot{In the present case this is 
 \rf{loi} combined with the Weyl transformation 
$g'_{\m\n}(x) =  \xi^{-2} g_{\m\n}$  so that  $\td g'_{\m\n}(\td x) = 
\xi^{-2} {\del x^\a \ov\del  \td x^\m}{\del x^\b \ov \del \td x^\n} g_{\a\b}(x) = g_{\m\n}(\td x)$,
where $g_{mn}$ here stands for the pp-wave metric in (\ref{ppbdy}).
The standard Weyl symmetry transformation rules for a 4d scalar, spinor and a vector  are
$\phi'(x)= \xi \phi, \ \psi'(x)= \xi^{3/2} \psi, \ A'_\m(x) = A_\m (x), \ 
F'_{\m \n}(x) = F_{\m\n}(x)$ (then $F'^{\m \n}(x) = \xi^4 F^{\m\n}(x)$).}
Thus under this combined symmetry the Wilson line factor is invariant while 
the gauge field strength  transforms as 
$\tilde{F}_{\mu\nu}(\tilde{x}) = \frac{\partial x^\alpha}{\partial \tilde{x}^\mu}\frac{\partial x^\beta}{\partial \tilde{x}^\nu} F_{\alpha\beta}(x)$, i.e. 
%
\beq
 \tilde F_{+-}(\tilde{x}) = \xi^{-2} F_{+-}(x)\ , \ \ \ \ \
 \tilde F_{+i}(\tilde{x})=   {\xi}^{-1}   F_{+i}(x)\ , \ \ \ \ 
 \tilde F_{-i}(\tilde{x})=   \xi^{-3} F_{-i}(x)\ , \la{ooq}
\eeq
 Then the relevant components of the energy momentum tensor  scale as  
\beq
 \tilde T^{+}{}_+(\tilde{x})= \xi^{-4} T^+{}_{+}(x)\ , \ \ \ \ \ \ 
 \tilde T^{+}{}_-(\tilde{x})= \xi^{-6} T^+{}_{-}(x)\ .  \la{jwq}
\eeq
Thus the expectation values of their integrals in the state defined by the 
light-like Wilson line (which preserves the scale invariance) 
in \rf{jdq} 
should  formally scale as (the volume element  $dx_-d^2x_i$ scales as $\xi^4$)
\be \la{sc}
\td P_+ = \xi^0 P_+ \ , \ \ \ \ \ \ \td P_- =   \xi^{-2}  P_- \ . \la{qw} \ee
In practice, as we have seen above, {\it for  these  particular observables} 
defined in \rf{jdq} 
 the  scale invariance is broken by the UV
 cutoff, 
and \rf{qw} implies that $P_+$ 
 in \rf{jmq} should be  logarithmically divergent whereas $P_-$ should be 
quadratically divergent. 

Indeed, notice that the classical  gauge field background \rf{jssq} describing the semiclassical state
defined by the Wilson loop is indeed scale-invariant, i.e.  $\td F_{\m \n} (x) = F_{\m\n}(x) $ or 
$ F_{+-} (\td x) = \xi^{-2} F_{+-}(x), \   F_{+i} (\td x) = \xi^{-1} F_{+i}(x), 
 \ F_{-i} (\td x) = \xi^{-3} F_{-i}(x) $
 and thus  $
T^{+}_+(\tilde{x})= \xi^{-4} T^+{}_{+}(x) , \  
  T^{+}_-(\tilde{x})= \xi^{-6} T^+{}_{-}(x)$.
  The  same will,    in general,  apply   to the expectation values defined by path integrals over the gauge fields, e.g., to the 
   integrands   in \rf{jdq}, i.e.
\be 
\TT^{+}_+(\tilde{x})= \xi^{-4} \ \TT^+{}_{+}(x) , \  \ \ \ 
  \TT^{+}_-(\tilde{x})= \xi^{-6} \ \TT^+{}_{-}(x)   , \ \ \ 
    \TT^+_\pm \equiv  \langle\ T^+_\pm\ W\  \rangle_{_{\text{pp-wave}}}
    \ . \la{uuut}\ee 
 Since these relations determine the dependence of 
 $\TT^+_\pm$ on the relevant radial direction in 3-space $(x_-,x_1,x_2)$, namely\foot{Using angular coordinates we can  write the volume element as 
 $dx_- dx_1 dx_2 = dx_-  r dr d \phi =  \ha {\rm r} d {\rm r}  d \psi d \phi $, 
 where $x_- = {\rm r} \cos \psi, \ r^2 = {\rm r}  \sin \psi$.}
 ${\rm r}= \sqrt{r^4 + x_-^2}$, this implies that the integrals  giving $P_+$ and $P_-$
 will diverge, respectively,  logaritmically  and  quadratically 
 at  ${\rm r}\to 0$.
  Therefore, these  divergent parts 
can be related as  $P_+ \sim \ln |P_-|$. 


The   relation \rf{jyy} 
 following from the scaling symmetry 
implies also that  $E-S \sim \ln S$. 
This   shows once again that the logarithmic behavior 
of the anomalous dimension for large spin is a consequence of the 
symmetries as
was recently emphasized also  in \cite{am2}.  

Although this  1-loop check and the scaling argument 
are  encouraging it still remains to be seen 
 how  to extend the above computation  to higher loop 
orders.\footnote{In this case one should be careful in defining the vacuum such
 that the classical expectation values are as they were computed above.}

One may wonder if  the fact that the pp-wave background \rf{ppbdy} is conformally-flat 
(see \rf{kql},\rf{ql})  
\beq\la{lkl}
 2 dx_+ dx_- - \m^2 x^2_i dx_+^2  + dx_i^2 = 
 {1 \ov 1 + \mu^2 \ty_+^2}  (2 d\ty_+ d\ty_- + d\ty_i^2)\ ,  
\eeq
and that the SYM theory is conformally invariant 
implies that 
there should be an alternative way   to do the  above computation by staying 
directly in flat space.  The non-trivial relation \rf{igg}   between 
the $P_+$ and $P_-$ may then   come  out of  the anomaly due to
the presence of the UV cutoff that spoils the formal equivalence 
between the null  Wilson line in the  pp-wave background and its image 
in flat coordinates.

\section{Conclusions}

 In this paper we  have considered the large spin limit of the spiky strings
 in which they end at the boundary of $AdS_5$   and capture 
 the $\log S$ asymptotics of the energy. 
 
  After concluding that all solutions,
corresponding to different number of spikes, contain
 the same information about the cusp anomaly,
namely $f(\lambda)$, we studied the case of an infinite number of spikes.
In that limit, the solution simplified and effectively  described a 
string moving in a  5d  ``AdS  with pp-wave'' background.
Further analysis led  us to an ``elementary'' open string solution 
that ends at one light-like line at the boundary and contains all the necessary information 
required  to determine $f(\lambda)$.


 The dual field theory  turns out to be defined  in a corresponding  4d 
pp-wave space. From that point of view, the essential part of the
calculation  that determined $f(\lambda)$ happened  to be the computation 
of a logarithmic UV divergence in the integral of the energy momentum tensor  which gives 
the total momentum in the light-like direction.

 This led us to define a pp-wave anomaly as the coefficient
of such divergence. Using AdS/CFT we derive that, for the \N{4} SYM theory, the pp-wave anomaly 
coincides with the cusp anomaly,  a fact that we checked
 at lowest order in SYM perturbation theory. 
It would certainly be interesting to extend this computation to higher orders in $\l$ but that is beyond the 
scope of this paper.
It is also important to point out that, similarly to what was found in \cite{am2}, in this approach, 
the logarithmic growth of the anomalous dimension of twist two operators for large spin is a consequence
of the symmetries of the theory. 

 A simple way to understand why we  get the field theory in a pp-wave is that when the tip of the rotating string
touches the boundary it represents a field theory probe moving at the speed of light on $S^3$. The region near
the tip of the string, which is relevant in the large spin limit, corresponds, in the boundary, to the region near 
such light-like trajectory. But the metric near a light-like trajectory is described by the Penrose limit of the
$R\times S^3$ metric which is precisely the pp-wave metric. Therefore, from this point of view, the natural setup 
to study large spin operators is the field theory living in a pp-wave background.

 To arrive at the bulk  pp-wave on \ads{5} background \rf{ppbkg} we followed a procedure similar to 
that of \cite{pen,rev}, namely taking a 
limit in the metric, in this case of \ads{5}. By analogy to what was done in \cite{spinch} for the rotation on $S^5$, 
it might be useful to consider the same limit directly in the string action. 
%
%
The result should be of more general nature than our present discussion 
  and may help in understanding better  the AdS/CFT 
correspondence in the sector states represented by  strings moving in \ads{5}
(cf. \ci{bgk,sak}).

Finally, let us note that 
using the fact that the 4d pp-wave is conformally flat 
(and thus  \rf{ppbkg}  is locally equivalent to $AdS_5$)
    we are  able
 to find new string solutions in the standard $AdS_5$  space in Poincare coordinates,
 namely, the ones that 
end on a single light-like line, or on two parallel light-like  lines.
 If one considers two parallel light-like Wilson lines in flat 
 space, there is no dual string solution 
that respects the translation symmetry along  the light-like direction. 
The pp-wave can perhaps be  thought of 
as a  minimal modification to the metric that allows for such a solution.
If translation invariance is not
imposed,  then  one can find such a  solution in the usual Poincare patch as we show  in Appendix A. 
It would be interesting to analyze this type of solutions further
and  clarify their dual field-theory interpretation.

\begin{acknowledgments}

We are  grateful to R. Roiban for useful comments on a draft of this paper. 
The work of M.K. was supported in part by the National Science Foundation
under grant  No. PHY-0653357.
 A.A.T. acknowledges  the support of
the STFC, INTAS 05-1000008-7928
  and EC MRTN-CT-2004-005104  grants.
   
\end{acknowledgments}

\renewcommand{\theequation}{A.\arabic{equation}}
\renewcommand{\thesection}{A}
 \setcounter{equation}{0}
\setcounter{section}{1} \setcounter{subsection}{0}

\section*{Appendix A: Transformation from  ``\ads{5} with pp-wave''  \\
 back to \ads{5}}

 As was  mentioned  above,  the field theory in the  pp-wave \rf{ppbdy}
 should be  dual to string 
 theory in the  \ads{5} pp-wave background \rf{ppbkg}. It turns out 
that the pp-wave \rf{ppbdy} is conformally flat and, as a result,  the
 \ads{5} with  pp-wave background  \rf{ppbkg} is, locally, the same as the 
 \ads{5} space written  in different coordinates \cite{cha}.
 This is not totally surprising, given that \rf{ppbkg}  was obtained
 as a limit \rf{zc} of the standard  \ads{5}    metric in global coordinates. 
 
 To demonstrate this equivalence  let us  note that if we define
\beqa
 \td x_+ = \mu^{-1} \tan \m x_+ \ , \ \ \ \  
 \td x_- = x_- - \ha \mu  x_i^2 \tan \m x_+\ ,  \ \ \ \ \ 
 \td x_i = \frac{1}{\cos \m x_+} x_i \ , \la{kql}
\eeqa
then 
\beq\la{ql}
 2 d\ty_+ d\ty_- + d\ty_i^2 = \frac{1}{\cos^2 \m x_+} \left( 2 dx_+ dx_- - \m^2 x^2_i dx_+^2
  + dx_i^2\right) \ . 
\eeq
A similar  transformation
 that maps the special ``\ads{5} with  pp-wave'' 
   background \rf{ppbkg} into the Poincare patch of the 
 $AdS_5$ is  \cite{cha}
\beqa
 \ty_+ &=& \mu^{-1}\tan \m x_+ \ , \ \ \ \ \ \ 
 \ty_- =  x_- - \ha \mu  ( x_i^2 + z^2)  \tan \m x_+ \no \\
 \ty_i &=& \frac{1}{\cos \m x_+} x_i \ , \ \ \ \ \ 
 \tilde{z}  = \frac{1}{\cos \m x_+} z \ , \la{lol}
\eeqa
i.e. 
\beq
 \frac{1}{\tilde z^2}\left( 2 d\ty_+ d\ty_-  + d\ty_i^2  + d\tilde z^2 \right) = \frac{1}{{z}^2} 
 \left[2 dx_+ dx_-  - \m^2 (x_i^2 + {z}^2 ) dx_+^2  + dx_i^2 + d{z}^2\right]  \ . \la{poi}
\eeq
 Notice  that the  transformation \rf{kql}  is singular
  at $\m x_+=(2k+1)\frac{\pi}{2}$ , 
 $k\in\mathbb{Z}$, i.e.  it  maps
 part of the pp-wave  into the full flat space. Same applies to \rf{lol}.
  Therefore, 
  these pairs of spaces are  equivalent only
 locally.\foot{As an aside comment, let us mention that 
the minus sign we have in the pp-wave metric \rf{ppbkg} 
 in front of $dx_+^2$ is crucial for the above conclusions:
   for the  opposite 
sign the map involves hyperbolic instead of trigonometric 
functions and then we find that  part of the 
flat space is mapped  into the pp-wave.}  

\begin{figure}
\epsfig{file=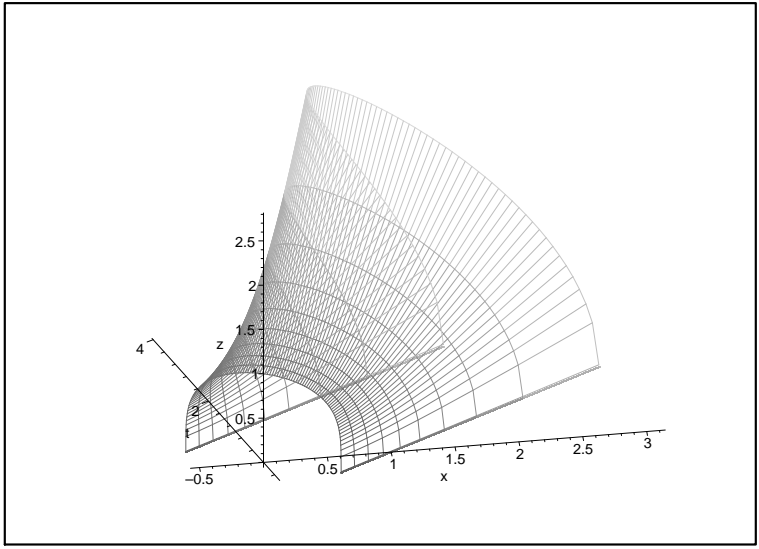, height=10cm, angle=270}
\caption{Surface ending on two parallel light-like Wilson lines given by $t=x\pm1$ in
 \ads{5} space in Poincare coordinates.}
\label{PWL} 
\end{figure}

 An interesting consequence of the existence of this  transformation 
  is that we can  map  the world-sheet surface
  ending on two parallel null lines  \rf{bas},\rf{zp}  
   we found  in the  pp-wave background  \rf{ppbkg}
 into a similar surface in the Poincare  patch of \ads{5}.
 The resulting  surface   still ends on two parallel light-like  lines at the boundary 
and is given by (we set $\m=1$)
\beqa
\ty_+ &=& \tan \tau  \ ,  \ \ \ \ \ \   \ \ \ \ \ \ \ty_i=0 
\no\\
\ty_- &=& \sigma - (\s_0^2 -\s^2)^{\ha}\, \tan\tau 
\no\\
\td z   &=& \frac{\sqrt{2}}{\cos\tau} \left(\s_0^2-\s^2\right)^{\frac{1}{4}}
 \ , \la{qqq}
\eeqa 
where $-\frac{\pi}{2}<\tau<\frac{\pi}{2},  \  -\s_0<\s<\s_0$.
A plot of this solution is given  in fig.\ref{PWL}.


\begin{figure}
\epsfig{file=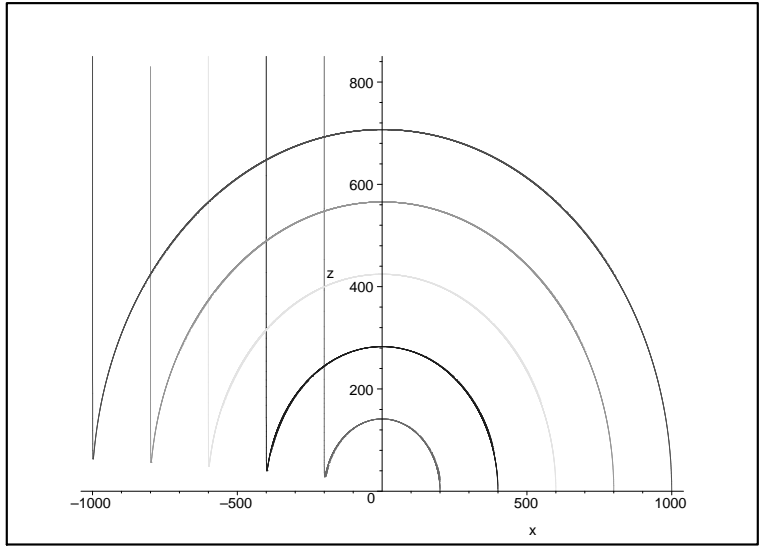, height=10cm, angle=270}
\caption{Surface ending on a single light-like line given by $x=t$. 
Here we plot the shape of the string $z(x)$ at different values of $t=200$, $400$, 
$600$, $800$, $1000$. The $x$-axis is the boundary where the string ends at $x=t$ (allowing to identify each curve), 
and  its  end-point we identify as a quark.
 In addition,  there is a (rounded) spike coming out of 
  the horizon at a point given by $x\simeq -t$ (at $x=-t$ we have $z=\infty$). The shape of the string 
at $t<0$ follows from the symmetry $(x,t)\rightarrow(-x,-t)$. Thus, the quark and the spike come together at $t<0$; at  around $t=0$ the 
spike actually disappears and then reappears again on the other side
 moving away from the quark as shown in the figure.
 If we associate the spike coming out of  the horizon with a gluon
  then this describes quark-gluon scattering but further analysis is needed in order to substantiate this 
  interpretation.}
\label{SWL}
\end{figure}

Similarly, we can find a surface that is the analog of \rf{sWL}, i.e. 
the one that ends on a single null line in the Poincare patch of \ads{5} (here $\s > 0$)
\beqa
 \ty_+ &=& \tan \tau \no \\
 \ty_- &=& -\ha \sigma^2 \tan \tau\no \\
 \tilde z   &=& \frac{\sigma}{\cos\tau} \ . 
  \la{pqq}
\eeqa
This implies that 
\beq
\td z =  \sqrt{-\frac{2\td x_-(1+\td x_+^2)}{\td x_+}} \ , \la{iqk}
\eeq
where the range of $\td x_\pm$ is restricted by $\td x_+\td  x_-<0$. This solution is depicted in fig.\ref{SWL}. The interpretation of
such solution is interesting to consider but appears unrelated to the main idea of this paper so we leave it for future work.

Let us mention that a  similar (euclidean) surface  appeared  in \cite{m} 
 (it corresponds to the case of  $\omega=1$ in eq.(11) of \ci{m}):\foot{The two surfaces 
 \rf{iqk}  and \rf{hf}
 are  formally related by a complex boost: $x_+ \to i \td x_+ , \ 
 x_- \to -i \td x_-$.}
\beq
z =\sqrt{\frac{2x_-(1-x_+^2)}{x_+}} \ . \la{hf}
\eeq
The surface \rf{hf}  was shown in \cite{m}  to correspond 
to  the straight infinite rotating string solution in $AdS_5$.
   This  appears to be   related to our  observation  of the 
  equivalence of  the arc between the spikes and the straight 
    string   by  an $SO(2,4)$  
  boost   made in the present paper.


\renewcommand{\theequation}{B.\arabic{equation}}
\renewcommand{\thesection}{B}
 \setcounter{equation}{0}
\setcounter{section}{1} \setcounter{subsection}{0}
 
\section*{Appendix B: Details of gauge  theory calculation in section 5}

To determine the  classical   field configuration 
which is sourced  by a null Wilson like, i.e. a  massless point-like 
 charge moving along $x_+$
direction in the pp-wave background \rf{ppbdy} 
 we need to solve the equations following from \rf{jqh} 
\beqa\la{ka}
&& \partial_{\mu} \left(\sqrt{-g} F^{\mu\nu}\right) = j^\nu \ , \\
&& j^+ = \gYM \delta(x_-)\delta^{(2)}(x_i) \ , \ \ \ \    \ \  \ \   j^-=j^i=0 \ . \la{jqa}
\eeqa
Introducing the vector potential through 
$F_{\mu\nu}=\partial_\mu A_\nu-\partial_\nu A_\mu$
and assuming that   on symmetry grounds ($i=1,2$) 
\be A_i=0 \ , \ \ \ \ \ \ \ 
A_\pm=A_\pm(r,x_-) \ , \ \ \ \  \ \ \ \ \ \ \ \   r^2\equiv x^2= x_1^2+x_2^2   \ , \la{kma}\ee
we get\foot{We  use that for the metric in \rf{ppbdy}
$g_{++}= -\m^2 x^2, \ g_{+-}=1 , \ g^{--}= \m^2 x^2, \ g^{+-}=1, \   g_{ij}=g^{ij}=\delta_{ij}$.} 
\beqa
  \partial_-^2 A_+  - \partial_i^2 A_- &=&  -\gYM \delta(x_-)\delta^{(2)}(x_i) \la{kka}\\
 -\partial_i^2 A_+ - 2 \m^2  x^i\partial_i A_- - \m^2 x^2 \partial_i^2 A_- &=& 0\la{kal} \\
 \partial_{i}\partial_{-} A_+ + \m^2  x^2 \partial_{i} \partial_{-} A_-&=& 0 \ . \la{kia}
\eeqa
The strategy that  we  will follow  is to consider first  the region $r\neq 0$ where 
the right hand side of  \rf{kka}  vanishes and thus we obtain a system of  homogeneous
equations. 

From \rf{kia}  we get ($r\equiv  \sqrt { x^2}$) 
\be  \del_r A_+ = -  \m^2 r^2 \del_r A_-= - \m^2  r^2 \chi  \ , \ \ \ \ \ \ \ 
\chi \equiv  \partial_r A_-   \  . \la{jl}
\ee
 Then,  differentiating  the first equation with 
respect to $r$  and eliminating $A_+$  using \rf{jl} we obtain
$ - \mu^2 r^2 \partial_-^2 \chi - \del_r (\del^2_r  + { 1 \ov r } \del_r)   A_-=0$, i.e. 
\beq
 - \partial_r^2 \chi  -\frac{1}{r} \partial_r \chi  + \frac{1}{r^2} \chi 
 - \mu^2 r^2 \partial_-^2 \chi = 0
 \ , \ \ \ \ \ \ \   \chi\equiv \partial_r A_- \ . \la{cha}
\eeq
After solving for $\chi$ we can compute $A_\pm$ through
\rf{jl}.\foot{At first sight, the integration over $r$ could   lead to 
  an arbitrary function of $r$  appearing  in $A_+$ 
 but the equations require it to vanish.}

 The equation for $\chi$ can be solved
 by  first  doing a Fourier transform in $x_-$  ($\chi(r,x_-)  \to  \td \chi(r,k_-)$)    
\beq
 -\partial_r^2 \td \chi - \frac{1}{r}\partial_r\td \chi  + ( \frac{1}{r^2} 
   + \mu^2  k_-^2 r^2) \td \chi=0\  . 
\eeq
 Integrating this  equation gives 
\beq
 \td \chi(k_-,r) = \frac{B}{r} e^{-\ha \m |k_-| r^2} \ , \la{jjg} 
\eeq
where we pick  the solution vanishing at $r\rightarrow \infty$. 
Here $B$ is a  dimensionless constant. 
 Fourier-transforming back in $k_-$  we find
\beq
\chi(x_-,r) = \frac{\ha B \mu  r}{\mu^2  r^4 +  4 x_-^2} \ . \la{gfg}
\eeq
This leads,   using  \rf{jl}, to 
\beqa
 A_- &=& \frac{B}{8 x_-} \arctan \frac{\m r^2}{2  x_-} \ ,  \la{qoi} \\
 A_+ &=& - \frac{B\mu }{8}   \ln(\mu^2 r^4+  4  x_-^2)  \ . \la{klu}
\eeqa
The corresponding  components of 
 $F_{\mu\nu}$ are 
\beqa
F^{+-}=-F_{+-} = -\frac{B \m  x_-}{\m^2  r^4 + 4 x_-^2}, \\
F_{+i}=\frac{ \ha B \m^3  r^2 x_i}{\m^2 r^4 + 4x_-^2}, \ \\
F^{+i}=F_{-i} = -\frac{\ha B \m x_i}{\m^2  r^4 + 4 x_-^2} \ . \la{oou}
\eeqa
Then 
$\partial_\mu F^{\mu-}=\partial_{\mu} F^{\mu i}=0$ 
and 
\beq
 \partial_{\mu} F^{\mu+} = \partial_- F^{-+} + \partial_i F^{i+} =  \gYM \delta(x_-)
  \delta^{(2)}(x_i)
 \  \la{klq}
\eeq
is also  satisfied   away from the origin in the $(x_1,x_2,x_-)$ space.

It remains to  fix the constant $B$ 
 by matching the   above  background onto  the source. 
 Like in the case of the usual static point-like   source this 
    can be done, for example,   by regularizing  the above expressions  \rf{oou}    for 
 $F_{\m\n}$ near the origin $r=0, \ x_-=0$. 
 Replacing there  
 \be \m^2  r^4 + 4 x_-^2\ \  \to\ \  \m^2  r^4 + 4 x_-^2 + \ve^4, \ \ \ \ \ \ \ \ \ \ 
  \ve \to 0 \ , \ee
 we find 
 \be \la{ree}
 \partial_- F^{-+} + \partial_i F^{i+} = { 2 B \m \ve^4   \ov (\m^2 r^4 + 4x_-^2 + \ve^4)^2}
 \ , \ \ \ \ \ \ \ \   \ve \to 0 \ .  \ee
 This indeed represents  the regularized   delta-function in the 3-space  $(x_1,x_2,x_-)$
 as required to match  the r.h.s. of \rf{klq}.
 Since 
 \be 
 2\pi \int^\infty_0  dr\ r  \int^\infty_{-\infty} dx_-\ 
  { 2 B \m \ve^4   \ov (\m^2 r^4 + 4x_-^2 + \ve^4)^2} = {\pi^2 \ov 2} B   \ , \ee
 we conclude that 
  \be  
B = \frac{2\gYM}{\pi^2}  \ .  \la{bee}
\ee
 Thus finally we find  from \rf{qoi},\rf{oou},\rf{bee}
   the 
 expressions \rf{bjq},\rf{jssq}  used in the main text.

Let us note  that the   scale   $\mu$ of the pp-wave  \rf{ppbdy}
 plays the role of a regularizing parameter for the above  solution. 
 Since  the  one-dimensional delta-function  has the representation 
 \be 
 \delta (y) = \frac{1 }{\pi}  \lim_{\mu\to 0}   \frac{\mu }{ y^2+\mu^2  } \ , 
 \ee
 we conclude that in  the flat-space limit $\mu\to 0$  
 the above solution becomes 
 \beqa
F^{+-}&=&  - \frac{2\gYM}{\pi } \  {x_- \ov r^4}  \ \delta
 ({ 2 x_- \ov r^2}) = - \frac{\gYM}{\pi } \  {x_- \ov r^2}  \ \delta
 ({  x_- }) =0      ,  \ \ \  
F_{+i}= 0    \no \\
F^{+i}&=&  
 - \frac{\gYM}{\pi }  \  {x_i \ov r^4} \  \delta ({ 2 x_- \ov r^2})   
 =  
 - \frac{\gYM}{2\pi }  \  {x_i \ov r^2} \  \delta ({ x_- })   
 \ . \la{oouj}
\eeqa
 These  expressions 
formally give  $T^+_+=0$.
 In this sense  switching on the pp-wave  background 
 provides a regularization of the problem of  finding the field 
 produced by a point-like charge
 moving with the speed of light.

\renewcommand{\theequation}{C.\arabic{equation}}
\renewcommand{\thesection}{C}
 \setcounter{equation}{0}
\setcounter{section}{1} \setcounter{subsection}{0}
 
\section*{Appendix C: Generalization of solution in section 2.2 \\
 to rotation in $S^5$}

 The conformal-gauge parametrization  in \rf{jo} is useful for generalizing the 
 above ``one-arc''  solution \rf{kp}--\rf{roo}
 to the case when the   string has also a momentum $J$ along a circle in 
 $S^5$. All we need to do is to generalize  first the straight
  string solution  (i.e. \rf{jp}) 
 and then apply the two boosts as  in \rf{ki}. The point is 
 that this generalization modifies 
 only the conformal gauge constraints,
  but  in the embedding  coordinate 
 parametrization
 in conformal gauge  both the sigma model equations and the constraints   are 
 covariant under the $SO(2,4)$ transformations.
 
The conformal-gauge solution for the straight infinite string stretching 
to the boundary of $AdS_5$ that
rotates both in $AdS_5$ and $S^5$  with spins $S$ and $J$
is given by   \ci{ftt}: 
\be  
&&  t= \k \tau,\ \  \ \r= \mu \s,\ \  \ \ \theta = \k \tau,\ \  \ \ \vp= \nu \tau\ ,\la{jk} \\
&&  \k^2 = \mu^2 + \nu^2 \ , \ \ \   \mu= {1 \ov \pi} \ln {S\ov \sql}  \gg 1 \ , 
 \ \ \   \nu ={ J \ov \sql} \ , \ \   \ \ \ell \equiv { \nu \ov \mu} ={\rm fixed} 
  \la{js}
  \ee
where we use the $AdS_5$ metric in  \rf{ada}   and  $\vp$ is an angle  in $S^5$.
The corresponding solution in the embedding coordinates is (cf. \rf{jp})
\be 
Z_5 + i Z_6 = \cosh \mu \s \  e^{i \k \tau} \ , \ \ \ \ \ \ \ 
 Z_1 + i Z_2 = \sinh \mu \s \  e^{i \k \tau} \ . \la{kq}
 \ee 
   We may then use double-boost transformation  \rf{jo} to construct the corresponding 
 solution with non-zero  $\r_0  $: 
 \beqa
Y_1 &=& \sinh \mu \sigma  \cos \k \tau\ \cosh \r_0 - \cosh \mu\sigma  \sin \k\tau \ \sinh \r_0 \no\\
Y_2 &=& -\sinh\mu \sigma  \sin\k \tau\ \cosh \r_0 + \cosh\mu \sigma  \cos \k\tau \  \sinh \r_0 \no\\
Y_5 &=&  \cosh\mu \sigma  \sin\k \tau \ \cosh \r_0 +  \sinh \mu\sigma  \cos \k\tau\ \sinh \r_0\no \\
Y_6 &=& \cosh\mu \sigma  \cos\k \tau\  \cosh \r_0  -  \sinh\mu \sigma  \sin\k \tau \ \sinh \r_0  \la{jos}
\eeqa
 It is straightforward now to construct explicitly the 
 corresponding expressions  for $t',\r',\theta'$   coordinates 
 defined by 
 \be 
 Y_5 + i Y_6 = \cosh \r' \  e^{i t'} \ , \ \ \ \ \ \ \ 
 Y_1 + i Y_2 = \sinh \r' \  e^{i \theta'} \ ,  \la{jq}
 \ee 
  and find the  energy and spin  generators
  and a  relation between them.

 Note that  the generalization of eqs. \rf{ac} and \rf{op} 
 to non-zero $S^5$ rotation parameter $\nu$ are 
 \beqa
&& I = -T \int d\s d\t\  \sqrt{\r'{}^2 (\cosh^2\r -\omega^2\sinh^2\r - \nu^2) +
  \sinh^2\r \ (\cosh^2\r - \nu^2) } ,  \no \\
&&   {  \sinh^2 \r\ (\cosh^2 \r - \nu^2)  
    \ov \sqrt{ \r'^2 \
  (\cosh^2\r -\omega^2\sinh^2\r  - \nu^2)    + 
  \sinh^2 \r\  (\cosh^2\r - \nu^2)    }} \no   \\
 &&  \ \ \ \ \ \ \ \ \ \ \  =   \sinh \r_0\  \sqrt{   \cosh^2 \r_0 - \nu^2}   \la{opx}   \ ,
 \eea 
  where the constant of integration $\r_0$ is again the minimal value of $\r=\r_{\rm min}$. 
The   maximal  value of $\r$  corresponds now to $\coth^2\r_{\rm max} = { \omega^2 - \nu^2 \ov 1- \nu^2 } $, 
it can become infinite, i.e. the spikes  can  reach the boundary, provided $\omega=1$, namely $\theta=t$. 
Notice that this is the case we are considering in eq.(\ref{jk}).


\end{document}